\documentclass[journal]{IEEEtran}
\usepackage[utf8]{inputenc}
\usepackage{graphicx}
\usepackage{subcaption}
\usepackage{cite}
\usepackage{amsmath,amssymb,multirow}
\usepackage{amsmath,amssymb}
\DeclareMathOperator{\E}{\mathbb{E}}

\newtheorem{prop}{Proposition}
\usepackage{amsmath,amssymb}
\usepackage[noend]{algpseudocode}
\usepackage[linesnumbered,ruled,vlined]{algorithm2e}
\SetKwInOut{Input}{Input}
\SetKwInOut{Output}{Output\,}

\usepackage{graphicx}
\usepackage{textcomp}
\usepackage{xcolor}
\usepackage{color,soul}
\usepackage{dirtytalk}

\usepackage{scalerel}
\usepackage{tikz}
\usetikzlibrary{svg.path}
\usepackage[ruled,vlined]{algorithm2e}
\usetikzlibrary{svg.path}

\definecolor{orcidlogocol}{HTML}{A6CE39}
\tikzset{
  orcidlogo/.pic={
    \fill[orcidlogocol] svg{M256,128c0,70.7-57.3,128-128,128C57.3,256,0,198.7,0,128C0,57.3,57.3,0,128,0C198.7,0,256,57.3,256,128z};
    \fill[white] svg{M86.3,186.2H70.9V79.1h15.4v48.4V186.2z}
                 svg{M108.9,79.1h41.6c39.6,0,57,28.3,57,53.6c0,27.5-21.5,53.6-56.8,53.6h-41.8V79.1z M124.3,172.4h24.5c34.9,0,42.9-26.5,42.9-39.7c0-21.5-13.7-39.7-43.7-39.7h-23.7V172.4z}
                 svg{M88.7,56.8c0,5.5-4.5,10.1-10.1,10.1c-5.6,0-10.1-4.6-10.1-10.1c0-5.6,4.5-10.1,10.1-10.1C84.2,46.7,88.7,51.3,88.7,56.8z};
  }
}
\newcommand\orcidicon[1]{\href{https://orcid.org/#1}{\mbox{\scalerel*{
\begin{tikzpicture}[yscale=-1,transform shape]
\pic{orcidlogo};
\end{tikzpicture}
}{|}}}}

\usepackage{hyperref} 

\title{Mixture of Spectral Generative Adversarial Networks for Imbalanced Hyperspectral Image Classification}

\author{\IEEEauthorblockN{Tanmoy Dam~$^{\orcidicon{0000-0003-3022-0971}}$,
Sreenatha G. Anavatti~$^{\orcidicon{; 0000-0002-4754-8191}}$,
Hussein A. Abbass,~\IEEEmembership{Fellow,~IEEE}~$^{\orcidicon{0000-0002-8837-0748}}$}\\
\IEEEauthorblockA{School of Engineering and Information Technology, University of New South Wales Canberra,  Australia.}}

\begin{document}
\maketitle
\section{Abstract}
We propose a three-player spectral generative adversarial network (GAN) architecture to afford GAN with the ability to manage minority classes under imbalance conditions. A class-dependent mixture generator spectral GAN (MGSGAN) has been developed to force generated samples remain within the domain of the actual distribution of the data. MGSGAN is able to generate minority classes even when the imbalance ratio of majority to minority classes is high. A classifier based on lower features is adopted with a sequential discriminator to form a three-player GAN game. The generator networks perform data augmentation to improve the classifier\textquoteright s performance. The proposed method has been validated through two hyperspectral images datasets and compared with state-of-the-art methods under two class-imbalance settings corresponding to real data distributions.

{\bf Keywords:} Mixture Generators Spectral Generative Adversarial Networks, Class Imbalance.

\section{Introduction}
With the advances made in imaging spectrometer over recent decades, the hyperspectral image classification (HIC) problem has attracted significant attention by the research community~\cite{chang2007hyperspectral}. Due to the high-resolution continual bands, a spectrometer helps to capture the robust pixelwise information in images~\cite{zhang2018unsupervised}. Traditional machine learning algorithms are less effective in extracting the most enriching features from large hyperspectral images~\cite{feng2019classification}. Three broad learning techniques~\cite{feng2019classification} are applied to HIC: supervised, unsupervised and semi-supervised learning. The k-nearest neighbors (KNN), and support vector machine (SVM) algorithms are known to give robust performance in HIC~\cite{feng2019classification}. Recently, Convolutional Neural Networks (CNN) became the most widely used methods in the supervised deep learning domain~\cite{chen2016deep}. The performance quality of CNN is dependent on well distributed large scale labelled data. However, creating a large amount of labelled information for HIC is a costly and time consuming process.

In class-imbalanced problems, the performance of CNN  significantly deteriorates due to a tendency to bias the classifiers\textquoteright \ parameters towards the majority classes. In contrast, unsupervised learning algorithms don\textquoteright t require class information; rather they have been used to discover the number of classes present in a dataset~\cite{zhang2018unsupervised}. Semi-supervised algorithms take the few known labelled information as well as unlabelled information to improve the HIC performance for class-imbalanced dataset compared to unsupervised methods~\cite{zhang2017unsupervised}. The semi-supervised learning algorithms are broadly described in the literature as generative and discriminative methods~\cite{zhan2017semisupervised, li2009semi,  marconcini2009composite, cao2017hyperspectral}. Semi-supervised joint dictionary learning with soft-max($S^2JDL-Sof$) loss can be found in~\cite{zhan2017semisupervised}.

Recently, GAN has become a popular approach, where both the generative and discriminative neural networks rely on min-max game theoretic principles~\cite{goodfellow2014generative}. Due to the adversarial nature of learning GAN games, a discriminative network ($D$) is able to extract more fine features from the data to improve classification performance~\cite{radford2015unsupervised}. Unsupervised GAN has a better ability to capture latent features than traditional supervised classification methods~\cite{radford2015unsupervised, springenberg2015unsupervised}. When the class distribution is imbalanced, biasing the discriminator towards majority classes equally biases the classification accuracy performance towards majority classes~\cite{douzas2018effective}, despite that generative networks take the noise distribution from the categorical Gaussian distribution.

Conditional GAN-based hyperspectral spatial-spectral image classification~\cite{zhu2018generative} learns all sub-tasks in parallel. In contrast, the two-player sequential semi-supervised hyper-spectral GAN (HSGAN)~\cite{zhan2017semisupervised} is a semi-supervised hyperspectral sequential discriminator-and-generator-based GAN. The Multiclass Spatial-Spectral Generative Adversarial Network (MSGAN)~\cite{feng2019classification} is a spatial-spectral joint-learning-based two players GAN method, where the conditional class probabilities have been used to generate a specific real distribution. However, to reduce down the bias towards the majority class, an MSGAN discriminator relies on equal weighting for the generated classes.
Therefore, the distribution of generated samples is different from the real class distribution. However, parallel structure features based discriminator gives better classification performance compared to sequential discriminators in remote sensing dataset~\cite{lin2017marta}. The main contributions of this letter are as follows.
\begin{itemize}
  \item We propose a novel mixture of generators spectral 1-d GAN (MGSGAN) structure for HIC. The MGSGAN generator is able to generate class-dependent real data.
  \item To improve the classification performance, a new parallel feature extraction structure has been introduced, replacing the classic sequential structure in classifier networks.
  \item The proposed MGSGAN approach has been validated through two datasets: Indian Pines and Pavia University. A statistical significance study is also incorporated to validate the proposed method.
\end{itemize}

 \begin{figure*} \includegraphics[keepaspectratio=true,scale=0.5]{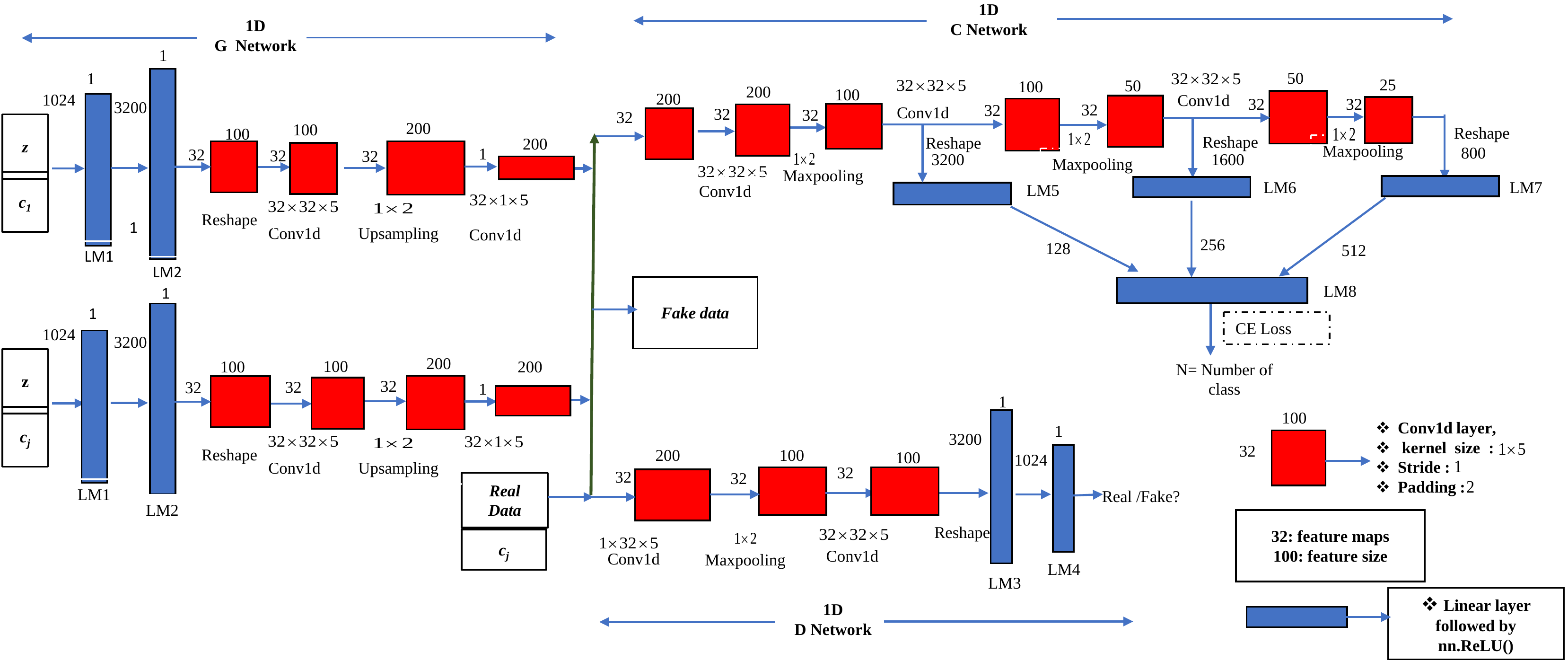}
 \caption{The 1-D MGSGAN architecture}\centering
 \label{MGSGANS_architecrture}
 \end{figure*}

\section{MGSGAN Structure}
In this section, we present the proposed MGSGAN and how the generators generate realistic images even in class imbalanced situations. When the real data distribution is imbalanced in nature, the conditional class distribution of the generating model can\textquoteright t generate the minority classes due to a bias towards majority classes in the discriminator networks~\cite{douzas2018effective}. To overcome this problem, the proposed MGSGAN employs a set of generators to generate a sample within the domain of specific real classes distribution. The proposed MGSGAN consists of three neural networks: a set of mixture of class conditionals generators ($G_j({\theta _g}) ={g_1, g_2, ..., g_N}$, $N$= number of classes present in the dataset), a discriminator ($D$) and a Classifier ($C$). The generator structure is represented as the sum of each class generator and its corresponding real class distribution. Thus class conditionals for each generator is defined as follows,
\begin{equation}
{G}_{j}(\frac{z}{c_j})=\sum_{j=1}^{N}g_j\tau_j
\label{eq:generate_samples}
\end{equation}
where, ${\tau_j} = { {{\tau_1} \cap {\tau_2} \cap ... \cap {\tau_N}}\in R^d}$ is a class specific domain information of real data distribution. Hence, the mixtures of generator outputs are the generated samples within the domain of each real class data distribution $(\tau_j)$. Due to domain constraints, MGSGAN can generate minority samples as well as majority samples. The $G_j(\theta_g)$) is differentiable CNN, parameterized by $\theta_g$, and takes the Gaussian normal distribution ($z\sim \mathcal{N}(-1,1)$) and conditional categorical class information ($c_j, j=1,...,N$) to generate the specific conditional class realistic distribution data ($p_r \in x_j$). The discriminator network ($D$), parameterized by $\theta_d$, acts to discriminate between real data ($p_r$) and generated realistic distribution data ($G_{j}(z/c_{j})$) as a real and fake classification, respectively. However, selecting the generator is a more crucial task, where we have used conditional categorical class information associated with conditional real data within the class$(\tau_j)$. Therefore, the generated samples should belong to a specific class of the real distribution domain. Moreover, the classifier network ($C$), parameterized by $\theta_c$, is  working as a normal classifier that can take class conditionals real data and generated realistic data. Hence, from the classifier perspective, the $G$ network is working as data augmentation network to generate realistic data for the majority and minority classes to improve the classification performance. However, $G$ and $D$ are working as normal two players GAN game. To learn a realistic data distribution, the differentiable $G$ network parameters $(\theta_g)$ learn by fooling the discriminator. In our proposed approach, all three network parameters are learnt jointly through min-max-max game principle.

The three players MGSGAN entropy loss objective function is defined as follows,
\begin{equation}
\label{eq:main}
\mathop {\min }\limits_G \mathop {\max }\limits_D \mathop {\max }\limits_C \,\,\, \,  Q(D,G,C) = L_D +L_G + L_C
\end{equation}
where,
\begin{equation}
\label{eq:main1}
L_D = P_j^r\E_{x \sim p_r} [logD(x)] + P_j^g \E_{G(\frac{z}{c_j}) \sim p_g}[log(1-D(G(\frac{z}{c_j})))]
\end{equation}
\begin{equation}
L_G = P_j^g\E_{G(\frac{z}{c_j}) \sim p_g}[log(1-D(G(\frac{z}{c_j})))] 
\label{eq:main2}
\end{equation}
\begin{equation}
L_C =  P_j^c \E_{x \sim p_r} [logC(x)] + P_j^c\E_{G(\frac{z}{c_j}) \sim p_g}[logC(G(\frac{z}{c_j})]
\label{eq:main3}
\end{equation}

Where $P_j^r$, $P_j^g$ and $P_j^c$ are the $j^{th}$ class conditional probabilities of real, generated and classifier data, respectively. In equation \ref{eq:main}, the $G$ network is only working with the $D$ network to reach stability, whereas the classifier network takes real data and augmented generated data to predict the correct class. Therefore, the optimal $D$ network is achieved by considering the $G$ network similar to the class conditionals GAN~\cite{goodfellow2014generative}. However, the classifier network performance is maximised based upon only the $G$ network. Thus, $G$ is working as class constraints data augmentation generator. Therefore, it becomes two players min-max game as in~\cite{goodfellow2014generative}. The optimal $D^*$ and $G^*$ can be derived by the following propositions.

\begin{prop}
For any $C$, the optimal discriminator($D^*$) is based on the fixed $G$ network. The optimal discriminator is defined as follows, 
\begin{equation}
\textbf(D)^*= \frac{P_j^rp_r}{ P_j^rp_r + P_j^gp_g}
\end{equation}
\end{prop}

\begin{prop}
The optimization of $Q(D,G^*,C)$ is the minimization of the following JS divergence. 
\begin{equation}
\textbf(G)^*= -2log2+ 2JS(P_j^r p_r ||P_j^g p_g)
\end{equation}
\end{prop}

The $C$ network gives maximum classification performance when $G$ and $D$ are reached at optimal points i.e. $p_r=p_g$. Hence, the class conditional $G$ network always generates a realistic sample within the domain of class conditionals real distribution $(p_r)$.

The $G$, $D$ and $C$ structures are depicted in Figure~\ref{MGSGANS_architecrture} where three networks are using the same convolutional kernels. The MGSGAN algorithm is described in Algorithm 1.

\begin{algorithm}[h]
\caption{MGSGAN Algorithm}
  \Input{training data ${\left\{ {{{x}_j \in R^d},{y_j }} \right\}^S}$, testing data ${\left\{ {{{x}_j \in R^d},{y_j}} \right\}^T}$, epochs=$1500$, batch=$64$, ADAM optimizer with $\beta_1 =0.5$ \& $\beta_2=0.999$, lr= $0.0002$ for $G, D \& C$  }
  \Output{$C$ output for testing dataset}
Begin\\
 Xavier normal initialization \cite{odena2017conditional} $\mathcal{N}(0,std^2)$ where,
$\text{std} = \text{gain} \times \sqrt{\frac{2}{\text{fan\_in} + \text{fan\_out}}}, \,\,\, gain=1$ \\
 
\For{each epoch}{
    samples from random Gaussian noise ${z \in R^{100}}$\\
    class conditionals one hot code ${c_j} \in {y_j \in R^N}$ \\
    \For{ samples from every $batch$}{
    concatenate random noise $\textbf{z}$ with $c_j$\\
    generate labels samples $c_j$ by using Eq. \ref{eq:generate_samples}\\
    Update $D$ networks by using Eq.\ref{eq:main1}.\\
        while $C$ \& $G_{j}(...)$ networks are fixed \\
    Update $C$ networks by using Eq.\ref{eq:main3}\\
       while $D $\& $G_{j}(...)$ networks are fixed \\
    Update $G_{j}(...)$ networks by using Eq.\ref{eq:main2}\\
       while $C \& D$ networks are fixed 
      }
    }
\label{algorithm:MGSGANs}
\end{algorithm}

\section{results \& Experiments}
In this section, we compare the performance of the proposed MGSGAN method along with other conditional GAN methods and two popular machine learning algorithms. The performance has been validated through two popular imbalanced hyperspectral dataset under two different training to testing ratio (TTTR) settings. 

\begin{table*}[]
\caption{Classification Performance on Indian Pines Dataset }
\resizebox{1\textwidth}{!}{%
\begin{tabular}{llllllllll}
\hline
Samples               & P.I       & SVM                           & KNN                            & CNN                           & ACHSGAN      & HSGAN & $S^{2}$JDL-Sof                & ACSGAN                         & MGSGAN                        \\ \hline
\multirow{ 3}{*}{${5\%}$} & OA($\%$)    & $68.46 \pm 1.05$ & $68.32 \pm 0.56$  & $73.13 \pm 1.56$ & $71.26 \pm 1.05$ & $74.92 \pm 0.41$ & - &  $77.469 \pm 0.26$ & $  \textbf{81.29} \pm \textbf{0.477}$ \\
                      & Kappa($\%$) & $64.33\pm 1.14$ & $63.81 \pm 0.657$ & $69.31 \pm 1.76$ & $67.48 \pm 1.45$ & $72.00 \pm 0.01$ & - & $75.69 \pm 1.45$  & $\textbf{78.64}\pm \textbf{0.57} $ \\
                      & AA($\%$)    & $62.27 \pm 2.48$ & $55.61 \pm 0.98$  & $65.21\pm 2.5$  & $64.26 \pm .97$ &  $70.97 \pm 0.55$  & -  & $73.40 \pm 2.89$  & $\textbf{77.05}\pm \textbf{2.78}$\\ \hline
\multirow{ 3}{*}{${10\%}$} & OA($\%$) &  $73.55 \pm 0.49$ & $73.361 \pm 0.59 $ & $82.12 \pm 0.36$ & $79.63  \pm 0.69$ &  $83.53 \pm 0.87$ & $82.25 \pm 1.08$  & $84.94 \pm 0.469$ & $\textbf{86.16} \pm \textbf{0.63}$ \\
                      & Kappa ($\%$) & $70.14 \pm 0.53$ & $69.43 \pm 0.66$ & $79.58 \pm 1.22$ & $76.03 \pm 0.77$ &  $80.01 \pm 0.01$ & $79.01 \pm 0.01$ & $82.83 \pm 0.53$ & $\textbf{84.20} \pm \textbf{0.72}$ \\
                      & AA ($\%$) & $74.30 \pm 0.55$ & $67.48 \pm 1.26$ & $78.10 \pm 1.72$ &$75.3 \pm 2.1$ & $79.27 \pm 0.60$ & $63.51 \pm 0 .66$  & $82.24 \pm 1.23$ & $\textbf{85.23} \pm \textbf{1.089}$ \\ \hline                          
\end{tabular}}
\label{overall_classification_performance}
\end{table*}
\begin{table*}[]
\caption{Each Class Classification Performance on Indian Pines Dataset}
\resizebox{1\textwidth}{!}{
\begin{tabular}{lllllll}
\hline
\multicolumn{1}{l}{Method} & \multicolumn{1}{l}{SVM} & \multicolumn{1}{l}{KNN} & \multicolumn{1}{l}{CNN} & \multicolumn{1}{l}{ACHSGAN \cite{zhan2017semisupervised}} & \multicolumn{1}{l}{ACSGAN} & \multicolumn{1}{l}{MGSGAN} \\ \hline
Alfalfa & $58.04 \pm 9.98$ & $32.68 \pm 16.47$ & $71.95 \pm 10.97$ & $61.78 \pm 13.98$ & $82.11 \pm  4.14$ & $\textbf{83.00} \pm \textbf{5.52}$ \\
Corn-notill & $63.04 \pm 3.34$ & $60.41 \pm 1.71$ & $59.79 \pm 14.61$ & $64.51 \pm 5.16$  & $80.84 \pm 2.12$ & $\textbf{81.62} \pm \textbf{2.83}$ \\
Corn-min & $61.42 \pm 3.75$ & $55.48 \pm 2.15$ & $73.123 \pm 0.20$ & $60.45 \pm 6.20$ & $75.67  \pm  2.58$ & $\textbf{76.06} \pm \textbf{5.07}$ \\
Corn & $62.28 \pm 7.97$ & $31.12 \pm 5.47$ & $66.90 \pm 8.6$ & $54.92 \pm 3.77$ & $62.12 \pm 7.17$ & $\textbf{68.92} \pm \textbf{12.13}$ \\
Grass-pasture & $92.16 \pm 1.47$ & $84.390 \pm 1.63$ & $88.62 \pm 0.80$ & $82.22 \pm 4.88$ & $90.03 \pm 1.98$ & $\textbf{90.62} \pm \textbf{.8}$ \\
Grass-trees & $93.89 \pm 1.58$ & $93.56 \pm 1.21$ & $95.13 \pm 1.00$ & $95.64 \pm 1.54$ & $96.04 \pm 0.96$ & $\textbf{96.92} \pm \textbf{.37}$ \\
Grass-pasture-mowed & $78.4 \pm 13.41$ & $79.6 \pm 8.28$ & $75.00 \pm 5.7$ & $89.33 \pm 3.77$ & $76.00 \pm 8.64$ & $\textbf{90.2} \pm \textbf{4.75}$ \\
Hay-windrowed & $94.86 \pm 1.00$ & $92.65 \pm 0.98$ & $95.58 \pm 1.3$ & $95.364 \pm 1.05 $& $97.28 \pm 0.955$ & $\textbf{98.69} \pm \textbf{0.75}$ \\
Oats & $36.11 \pm 17.61$ & $15 \pm 11.92$ & $58.33 \pm 8.3$ & $57.40 \pm 20.45$ & $53.703 \pm 14.58 $& $\textbf{86.66} \pm \textbf{12.95}$ \\
Soybean-notill & $62.44 \pm 2.37$ & $75.59 \pm 3.68$ & $63.72 \pm 11.2$ & $71.54 \pm 3.079$  & $81.045 \pm 0.60$ & $\textbf{85.24} \pm \textbf{5.6}$ \\
Soybean-mintill & $62.85 \pm 2.41$ & $79.08 \pm 1.28$ & $82.91 \pm 1.4$ & $77.15\pm 2.83$ & $83.29 \pm 0.97$ & $\textbf{84.89} \pm \textbf{1.7}$ \\
Soybean-clean & $64.990 \pm 5.74$ & $45.02 \pm 3.25$ & $79.08 \pm 1.7$ & $70.168 \pm 3.62$  & $86.053 \pm 4.28$ & $\textbf{87.24} \pm \textbf{4.27}$ \\
Wheat &$ 94.06 \pm 2.74$ & $93.62 \pm 2.59$ & $94.48 \pm 1.8$ & $93.69 \pm 2.93$ & $95.099 \pm 0.26$ &$ \textbf{98.27} \pm \textbf{.92}$ \\
Woods & $92.05 \pm 2.63$ & $91.39 \pm 2.79$ & $93.54 \pm 1.3$ & $95.14 \pm 0.90$ & $96.48 \pm 0.68$ & $\textbf{96.90} \pm \textbf{1.44}$ \\
Buildings-Grass-Trees & $50.28 \pm 1.03$ & $31.66 \pm 1.88$ & $61.35 \pm 3.3$ & $57.18 \pm 1.078$ & $63.40 \pm 5.85$ & $\textbf{64.96} \pm \textbf{7.54}$ \\
Stone-Steel-Towers & $83.00 \pm .85$ & $84.40 \pm 2.63$ & $88.09 \pm 0.1$ & $87.69  \pm 1.48$  & $90.66 \pm 1.01$ & $\textbf{92.14} \pm \textbf{1.61}$ \\ \hline
OA & $73.55 \pm 0.49$ & $73.361 \pm 0.59 $ & $82.12 \pm 0.36$ & $79.63  \pm 0.69$ & $84.94 \pm 0.469$ & $\textbf{86.16} \pm \textbf{0.63}$ \\
Kappa & $70.14 \pm 0.53$ & $69.43 \pm 0.66$ & $79.58 \pm 1.22$ & $76.03 \pm 0.77$ & $82.83 \pm 0.53$ & $\textbf{84.20} \pm \textbf{0.72}$ \\
AA & $74.30 \pm 0.55$ & $67.48 \pm 1.26$ & $78.10 \pm 1.72$ & $75.3 \pm 2.1$ & $82.24 \pm 1.23$ & $\textbf{85.23} \pm \textbf{1.09} $ \\
Time (each epoch) & 0.1 sec & 0.1 sec & 48sec & 43 sec & 45sec & 58sec \\ \hline
\end{tabular}
}
\label{classification_performance}
\end{table*}


\begin{table*}[] 
\caption{Classification Performance on Pavia University Dataset}
\resizebox{1\textwidth}{!}{%
\begin{tabular}{llllllllll}
\hline
Samples               & P.I       & SVM                           & KNN                            & CNN                           & ACHSGAN      & HSGAN\cite{feng2019classification} & MSGAN-Spectral \cite{feng2019classification}                & ACSGAN                         & MGSGAN                        \\ \hline
\multirow{ 3}{*}{${1\%}$} & OA($\%$)    & $78.59 \pm 0.67$ & $78.20 \pm 0.44$  & $83.45 \pm 0.59$ & $81.76 \pm 0.46$ & - & - &  $86.68 \pm 1.2$ & $  \textbf{90.51} \pm \textbf{0.4}$ \\
                      & Kappa($\%$) & $76.21\pm 1.3$ & $72.10 \pm 0.81$ & $81.95 \pm 0.73$ & $80.16 \pm 1.05$ & - & - & $84.24 \pm 1.5$  & $\textbf{87.39}\pm \textbf{0.65} $ \\
                      & AA($\%$)    & $77.5 \pm 0.49$ & $73.02 \pm 1.8$  & $82.5\pm 1.3$  & $80.34 \pm 0.78$ &  -  & -  & $85.13 \pm 0.75$  & $\textbf{88.12}\pm \textbf{0.78}$\\ \hline
\multirow{ 3}{*}{${3\%}$} & OA($\%$) &  $83.42 \pm 1.01$ & $82.97 \pm 1.36 $ & $88.30 \pm 0.27$ & $85.3  \pm 0.22$ &  $85.7 \pm 0.5$ & $86.1 \pm 0.7$  & $90.410 \pm 0.480$ & $\textbf{92.70} \pm \textbf{0.32}$ \\
                      & Kappa ($\%$) & $79.75 \pm 1.46$ & $76.57 \pm 2.79$ & $85.06 \pm 0.65$ & $82.5 \pm 0.4$ &  $81.4 \pm 0.5$ & $82.9 \pm 1.5$ & $87.22 \pm 0.77$ & $\textbf{90.3} \pm \textbf{0.42}$ \\
                      & AA ($\%$) & $80.27 \pm 1.72$ & $79.46 \pm 1.5$ & $86.1 \pm .97$ & $83.2 \pm .67$ & $84.3 \pm 4.0$ & $81.5 \pm 0 .8$  & $89.25 \pm .34$ & $\textbf{91.07} \pm \textbf{0.1}$ \\ \hline                          
\end{tabular}}
\label{overall_classification_performance_pavia_dataset}
\end{table*}

\subsection{Indian Pines Dataset}

A vegetation site in northwestern Indiana, the data was captured by Airborne Visible/infrared Imaging Spectrometer Sensor in $1992$. The $220$ spectral bands ranging  from $400$ to $2500$ nm channels  were  used  for  collecting $145\times145$ pixels  data. Due to atmospheric  turbulence, $20$ channels  data  have been corrupted, leaving us with $200$ channels and $16$ different classes to evaluate MGSGAN \cite{feng2019classification, zhan2017semisupervised}. The implementation was done using PyTorch (GeForce RTX 2060, 6GB GPU Ram) and python environments.

Table~\ref{overall_classification_performance} reports the performance of the proposed MGSGAN in comparison to the conventional and state-of-the-art methods. Two training-to-testing ratios (TTTR) $(5\% \,\,and\,\,\, 10\%)$ of real data distributions test cases are used for validating the proposed MGSGAN performance with other two class conditional GAN models: the two players Auxiliary Classifier Hyperspectral Spectral GAN (ACHSGAN) and the three players Auxiliary Classifier Single GAN (ACSGAN).

As in ACHSGAN~\cite{odena2017conditional}, the $G$ and $D$ networks are both working as data generation and data classification, simultaneously to improve classification performance. Thus, $D$ sequential networks gives $N+1$ outputs where $N$ is true number of classes and last output is used for adversarial GAN objective. $G$ and $D$ networks are similar to HSGAN structure~\cite{zhan2017semisupervised} where $D$ networks is similar to traditional sequential convoluted structure~\cite{radford2015unsupervised}. In ACSGAN, we use  without- domain constraints class conditional single generator instead of multiple generators of MGSGAN structure and remaining two networks ($C$ \& $D$) are the same.  However, $C$ network is working as normal lower dimensional features based classifier and we have assigned it as a CNN in the Table~\ref{overall_classification_performance}. 

The classification performance is compared with state-of-the-art methods such as HSGAN and $S^{2}$JDL-Sof~\cite{zhan2017semisupervised}. Three popular performance indices (PIs) are used to check superiority among the methods, such as overall Accuracy (OA), kappa coefficients (kappa) and average accuracy (AA)~\cite{zhan2017semisupervised}.
$5\%$ samples have been used to train all six methods and remaining $95\%$ samples were used to test the classification performance. The classification performance of MGSGAN has improved $4.94\%$ in OA, $3.9\%$ in kappa and $4.97\%$ in AA compared to the ACSGAN. The significant improvement of classification performance is been obtained as $8.5\%$ in OA, $9.2\%$ in kappa and $8.56\%$ in AA by MGSGAN compared to the state-of-the-art HSGAN method ~\cite{zhan2017semisupervised}.

It is observed from Table~\ref{overall_classification_performance} that MGSGAN  obtained better performance among the six methods. 
For the second TTTR case, $(10\%)$  and ($90\%)$ samples were used for training and testing respectively. The improvement of performance indices are $1.5\%$ in OA, $1.65\%$in kappa and $3.63\%$ in AA compared to the ACSGAN. The notable improvement of classification performance achieves $5.23\%$ in OA, $3.19\%$ in kappa and $7.5\%$ in AA by MGSGAN compared to the HSGAN method, although, in HSGAN method \cite{zhan2017semisupervised}, all the samples were used to train the generative models without labels information. Once the training has been completed, $10\%$ labelled samples are used to further train the last discriminator layers. However, in HSGAN, the  OA is $78.6\%$ while $10\%$ samples were used to train the model \cite{zhan2017semisupervised}. The class conditional two players ACHSGAN achieved better results of $79.63\%$ in OA while using the the same $10\%$ samples for training in HSGAN. It is also seen that the significant performance improvement of adversarial MGSGAN is by $4.9 \%$ OA, $5.81 \%$ kappa, $9.12 \%$ AA as compared to CNN.

Table~\ref{classification_performance} represents each class accuracy of available classes present in the Indian pines dataset, where $10\%$ samples are used to train all the methods. Table~\ref{classification_performance} contains each class average accuracy and its corresponding standard deviation for six methods over 10 runs. In addition, the first $16$ rows are recorded information about each class accuracy and the last three rows define OA, kappa and AA for all the classes in Table~\ref{classification_performance}. 

\begin{figure}
\begin{subfigure}{.48\textwidth}
  \centering
\includegraphics[width=.48\linewidth]{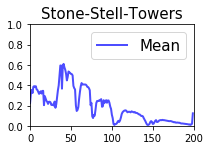} 
\includegraphics[width=.48\linewidth]{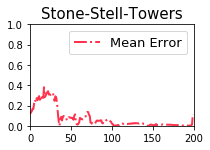}
  \caption{ACHSGAN output}
  \label{fig:ACHSGAN}
\end{subfigure}
\begin{subfigure}{.48\textwidth}
  \centering
  \includegraphics[width=.48\linewidth]{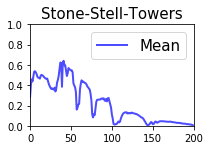} 
\includegraphics[width=.48\linewidth]{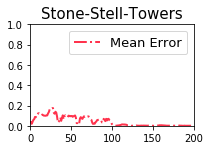} 
  \caption{ACSGAN output}
  \label{fig:ACSGAN}
\end{subfigure}

\begin{subfigure}{.48\textwidth}
  \centering
\includegraphics[width=.48\linewidth]{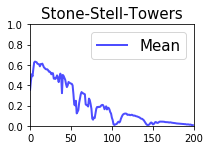}  
\includegraphics[width=.48\linewidth]{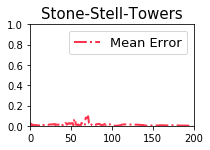}
  \caption{MGSGAN output}
  \label{fig:MGSGAN}
\end{subfigure}
\begin{subfigure}{.48\textwidth}
  \centering
\includegraphics[width=.48\linewidth]{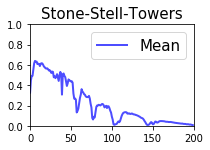}   \caption{Ground-Truth}
  \label{fig:ground_truth}
\end{subfigure}
\caption{Three generated model output for one minority class data}
\label{fig:generated_image_of_all_algorithms}
\end{figure}

Fig~\ref{fig:generated_image_of_all_algorithms} depicts one of the minority classes for the three GAN methods, where second TTTR conditions have been used for better visual illustration. Hyperspectral data contains a series of spectral bands of each class data. Therefore, we have plotted the average values of each classes over the number of spectral bands present in second TTTR training conditions.

Fig~\ref{fig:generated_image_of_all_algorithms} shows the average values of generated samples for `Stone-Steel-Towers' and its corresponding ground-truth. It is clearly observed that our proposed method MGSGAN has the ability to generate   `Stone-Steel-Towers' whereas the majority sample `Soyabean-mintill' intervene into domain of 'Stone-Steel-Towers' for ACHSGAN and ACSGAN methods. Due to better generation of minority classes and parallel structure of the $C$ network, the classification performance has improved significantly as observed in Table~\ref{overall_classification_performance}.

\subsection{Pavia Dataset}
The Pavia dataset was captured at Pavia University by Reflective optics image spectrometer sensors. The dataset contains nine classes with pixel sizes $610*340$. 115 bands were captured ranging from $430nm$ to $860 nm$ from where $12$ noisy bands have been removed during the data prepossessing stage. 
To compare again our proposed method, the two TTTR settings $(1\% \,\,and\,\,\, 3\%)$ have been considered similar to~\cite{feng2019classification}. The classification performance with other state-of-the-methods are listed in Table~\ref{overall_classification_performance_pavia_dataset}. For both settings, the proposed method has achieved better performance in terms of all the three index parameters. MSGAN-spectral method \cite{feng2019classification} is also considered along with other state-of-the-art methods. The classification performance of MGSGAN has improved $2.5\%$ in OA, $3.5\%$ in kappa and $2.03\%$ in AA compared to the second best results of ACSGAN. Similarly, for second TTTR case, the significant improvement of MGSGAN performance compared to the ACSGAN, in terms of three parameters are $4.44\%$ in OA, $3.7\%$ in kappa and $3.51\%$ in AA respectively.  It is observed from Table~\ref{overall_classification_performance_pavia_dataset} that the three player MGSGAN gives better performance than the two player ASHSGAN. 

We have also studied the statistical significance of our proposed method with other methods through McNemar\textquoteright s test ($M_t$) and the performance is shown in Table~\ref{mcnemar_test} for Indian pines and Pavia university datasets. The larger value of $M_t$ indicates statistical significance at a higher confidence. Similar to~\cite{zhang2018unsupervised}, we conclude that performance is statistically significant for $M_t>1.96 (5\% \alpha)$. 

\begin{table}[]
\caption{McNemar's Tests}
\resizebox{.5\textwidth}{!}{
\begin{tabular}{llllll}
\hline
MGSGAN & vs SVM & vs KNN & vs CNN & vs ACHSGAN & vs ACSGAN \\ \hline
Indian Pines        & 19.07  & 19.32  & 6.89   & 10.71       & 2.19  \\ \hline
Pavia University       & 33.42  & 32.17 & 17.10    & 26.73       & 9.44  \\\hline
\end{tabular}}

\label{mcnemar_test}
\end{table}

\vspace{-5 mm}
\section{conclusion}
In this letter, a novel mixture of spectral generator GAN has been proposed for generating minority and majority classes to improve classifiers\textquoteright \ performance for class-imbalanced datasets. To validate the proposed method, two train-to-testing ratio imbalance conditions for the Indian pines and Pavia University dataset have been considered. The proposed MGSGAN has improved classification performance compared to ACHSGAN, ACSGAN and spectral CNN. Our future work will consider mixtures of spatial-spectral GAN for improving the classification performance further. 

\bibliographystyle{IEEEtran}
\bibliography{reference}

\begin{thebibliography}{10}
\providecommand{\url}[1]{#1}
\csname url@samestyle\endcsname
\providecommand{\newblock}{\relax}
\providecommand{\bibinfo}[2]{#2}
\providecommand{\BIBentrySTDinterwordspacing}{\spaceskip=0pt\relax}
\providecommand{\BIBentryALTinterwordstretchfactor}{4}
\providecommand{\BIBentryALTinterwordspacing}{\spaceskip=\fontdimen2\font plus
\BIBentryALTinterwordstretchfactor\fontdimen3\font minus
  \fontdimen4\font\relax}
\providecommand{\BIBforeignlanguage}[2]{{%
\expandafter\ifx\csname l@#1\endcsname\relax
\typeout{** WARNING: IEEEtran.bst: No hyphenation pattern has been}%
\typeout{** loaded for the language `#1'. Using the pattern for}%
\typeout{** the default language instead.}%
\else
\language=\csname l@#1\endcsname
\fi
#2}}
\providecommand{\BIBdecl}{\relax}
\BIBdecl

\bibitem{chang2007hyperspectral}
C.-I. Chang, \emph{Hyperspectral data exploitation: theory and
  applications}.\hskip 1em plus 0.5em minus 0.4em\relax John Wiley \& Sons,
  2007.

\bibitem{zhang2018unsupervised}
M.~Zhang, M.~Gong, Y.~Mao, J.~Li, and Y.~Wu, ``Unsupervised feature extraction
  in hyperspectral images based on wasserstein generative adversarial
  network,'' \emph{IEEE Transactions on Geoscience and Remote Sensing},
  vol.~57, no.~5, pp. 2669--2688, 2018.

\bibitem{feng2019classification}
J.~Feng, H.~Yu, L.~Wang, X.~Cao, X.~Zhang, and L.~Jiao, ``Classification of
  hyperspectral images based on multiclass spatial--spectral generative
  adversarial networks,'' \emph{IEEE Transactions on Geoscience and Remote
  Sensing}, vol.~57, no.~8, pp. 5329--5343, 2019.

\bibitem{chen2016deep}
Y.~Chen, H.~Jiang, C.~Li, X.~Jia, and P.~Ghamisi, ``Deep feature extraction and
  classification of hyperspectral images based on convolutional neural
  networks,'' \emph{IEEE Transactions on Geoscience and Remote Sensing},
  vol.~54, no.~10, pp. 6232--6251, 2016.

\bibitem{zhang2017unsupervised}
M.~Zhang, J.~Ma, and M.~Gong, ``Unsupervised hyperspectral band selection by
  fuzzy clustering with particle swarm optimization,'' \emph{IEEE Geoscience
  and Remote Sensing Letters}, vol.~14, no.~5, pp. 773--777, 2017.

\bibitem{zhan2017semisupervised}
Y.~Zhan, D.~Hu, Y.~Wang, and X.~Yu, ``Semisupervised hyperspectral image
  classification based on generative adversarial networks,'' \emph{IEEE
  Geoscience and Remote Sensing Letters}, vol.~15, no.~2, pp. 212--216, 2017.

\bibitem{li2009semi}
J.~Li, J.~M. Bioucas-Dias, and A.~Plaza, ``Semi-supervised hyperspectral image
  classification based on a markov random field and sparse multinomial logistic
  regression,'' in \emph{2009 IEEE International Geoscience and Remote Sensing
  Symposium}, vol.~3.\hskip 1em plus 0.5em minus 0.4em\relax IEEE, 2009, pp.
  III--817.

\bibitem{marconcini2009composite}
M.~Marconcini, G.~Camps-Valls, and L.~Bruzzone, ``A composite semisupervised
  svm for classification of hyperspectral images,'' \emph{IEEE Geoscience and
  Remote Sensing Letters}, vol.~6, no.~2, pp. 234--238, 2009.

\bibitem{cao2017hyperspectral}
X.~Cao, C.~Wei, J.~Han, and L.~Jiao, ``Hyperspectral band selection using
  improved classification map,'' \emph{IEEE geoscience and remote sensing 3
  letters}, vol.~14, no.~11, pp. 2147--2151, 2017.

\bibitem{goodfellow2014generative}
I.~Goodfellow, J.~Pouget-Abadie, M.~Mirza, B.~Xu, D.~Warde-Farley, S.~Ozair,
  A.~Courville, and Y.~Bengio, ``Generative adversarial nets,'' in
  \emph{Advances in neural information processing systems}, 2014, pp.
  2672--2680.

\bibitem{radford2015unsupervised}
A.~Radford, L.~Metz, and S.~Chintala, ``Unsupervised representation learning
  with deep convolutional generative adversarial networks,'' \emph{arXiv
  preprint arXiv:1511.06434}, 2015.

\bibitem{springenberg2015unsupervised}
J.~T. Springenberg, ``Unsupervised and semi-supervised learning with
  categorical generative adversarial networks,'' \emph{arXiv preprint
  arXiv:1511.06390}, 2015.

\bibitem{douzas2018effective}
G.~Douzas and F.~Bacao, ``Effective data generation for imbalanced learning
  using conditional generative adversarial networks,'' \emph{Expert Systems
  with applications}, vol.~91, pp. 464--471, 2018.

\bibitem{zhu2018generative}
L.~Zhu, Y.~Chen, P.~Ghamisi, and J.~A. Benediktsson, ``Generative adversarial
  networks for hyperspectral image classification,'' \emph{IEEE Transactions on
  Geoscience and Remote Sensing}, vol.~56, no.~9, pp. 5046--5063, 2018.

\bibitem{lin2017marta}
D.~Lin, K.~Fu, Y.~Wang, G.~Xu, and X.~Sun, ``Marta gans: Unsupervised
  representation learning for remote sensing image classification,'' \emph{IEEE
  Geoscience and Remote Sensing Letters}, vol.~14, no.~11, pp. 2092--2096,
  2017.

\bibitem{odena2017conditional}
A.~Odena, C.~Olah, and J.~Shlens, ``Conditional image synthesis with auxiliary
  classifier gans,'' in \emph{Proceedings of the 34th International Conference
  on Machine Learning-Volume 70}.\hskip 1em plus 0.5em minus 0.4em\relax JMLR.
  org, 2017, pp. 2642--2651.

\end{thebibliography}

\end{document}